\documentclass[journal=nanoletters,manuscript=letters]{achemso}
\usepackage[version=3]{mhchem} 
\usepackage{graphicx}
\usepackage{dcolumn}
\usepackage{bm}
\usepackage{courier}
\usepackage{amsmath,amssymb}
\usepackage{mathrsfs}
\usepackage{bm}
\usepackage{supertabular}
\usepackage{float}
\usepackage{soul}
\usepackage{color}
\usepackage{natbib}
\usepackage{placeins}
\usepackage{array}
\usepackage{multirow}
\usepackage{afterpage}
\usepackage{lipsum}
\usepackage[utf8]{inputenc}
\usepackage{csquotes}
\usepackage{braket}
\usepackage{booktabs}
\usepackage{epstopdf}
\usepackage{comment}
\title{Tunable Carrier Dynamics in Carbide Antiperovskites via A-Site Cation Substitution}

\author{Sanchi Monga}
\email{sanchi@physics.iitd.ac.in[SM]}
\author{Saswata Bhattacharya}
\email{saswata@physics.iitd.ac.in [SB]}
\affiliation{Department of Physics, Indian Institute of Technology Delhi, New Delhi 110016, India}

\begin{document}

\begin{abstract}
	We present a comprehensive first-principles investigation of the electronic structure and excited-state carrier dynamics in the carbide antiperovskites Ca$_6$CSe$_4$ and Sr$_6$CSe$_4$. Using many-body perturbation theory ($\mathit{G_0W_0}$/BSE), we show that both materials are direct band gap semiconductors with quasiparticle gaps of 1.66~eV (Ca) and 1.22~eV (Sr), lying in the visible–near-infrared range, and exhibit moderate excitonic binding energies.
	\textit{Ab initio} nonadiabatic molecular dynamics simulations at 300~K reveal distinct relaxation mechanisms governed by the interplay of band gap, nonadiabatic (NA) couplings, and electronic decoherence. In Ca$_6$CSe$_4$, stronger lattice fluctuations induce $\sim$38\% larger band gap variations and $\sim$28\% faster decoherence, which, together with $\sim$53\% weaker NA couplings,
	suppress nonradiative recombination and yield lifetimes nearly 18 times longer (40.3~ns) than in Sr$_6$CSe$_4$ (2.2~ns).
	Hot-carrier cooling in both systems occurs on picosecond timescales (1–9~ps) with a pronounced slow down near the band egdes.
	Overall, our results demonstrate that A-site cation substitution provides an effective route to control carrier lifetimes and relaxation pathways in antiperovskites, offering microscopic insight into lattice-driven carrier dynamics and guiding their experimental realization and optimization.
\end{abstract}
\maketitle

The search for materials with tunable electronic structure and strong light-matter interactions has revitalized interest in antiperovskites—electronically inverted perovskites with the general formula $X_3BA$ ($X$ = cation; $A$, $B$ = anions of different sizes). Their structural flexibility and rich chemical tunability give rise to diverse emergent properties, positioning them as promising candidates for photovoltaics \cite{elseman2025antiperovskite,kalita2024anti}, thermoelectrics \cite{lin2024strong,hu2025prediction}, superconductivity \cite{li2025high,hoffmann2022superconductivity}, solid electrolytes \cite{xiang2025ionic,dutra2023computational,xia2022antiperovskite}, superionic conductors \cite{guan2024data}, and topological insulators \cite{bayaraa2024intrinsic}. 

Among these, inorganic nitride antiperovskites ($X_3$N$A$; $X$ = Mg, Ca, Sr, Ba; $A$ = P, As, Sb, Bi) have been widely explored for optoelectronic applications owing to their favorable band gaps, dispersive band edges, high carrier mobilities, and strong near-edge optical absorption \cite{monga2024theoretical,dai2019bi,kang2022antiperovskite,dahbi2022earth,sreedevi2022antiperovskite,farid2021optoelectronic,mochizuki2020theoretical}. Expanding the chemical space, Hahn and coworkers \cite{han2021design} proposed quaternary antiperovskites of the form $X_6B_2AA'$ and $X_6BB'A_2$, identifying 
several thermodynamically stable compounds as potential photovoltaic absorbers. More recently, imide-based defect antiperovskites $AE_5$As$Pn$(NH)$_2$ ($AE$ = Ca, Sr; $Pn$ = Sb, Bi) have further broadened the accessible bonding motifs and electronic landscapes in this family \cite{chau2025defect}.

In contrast, carbide-based antiperovskites remain relatively underexplored, particularly with regard to their excited-state and dynamical properties \cite{fang2024elucidating,zhang2024exploring,li2025unlocking,guo2024adjustable}. A notable advance by Guo \textit{et al.} \cite{guo2022anti} predicted a family of carbide antiperovskites $M_6$C$Ch_4$ ($M$ = Ca, Sr, Ba; $Ch$ = S, Se, Te) through ion inversion from the Cs$_4$Pb$X_6$--type structure. Among these, Ca$_6$CSe$_4$ and Sr$_6$CSe$_4$ stand out with suitable band gaps, high carrier mobilities, and strong visible-light absorption$-$comparable to the benchmark halide perovskite MAPbI$_3$. Their distinct bonding and lattice flexibility are expected to profoundly influence electron-phonon interactions and nonadiabatic dynamics, making them an ideal platform for studying carrier relaxation. Nevertheless, while their ground-state electronic structure has been characterized, key aspects$-$including quasiparticle corrections, optical response, and especially ultrafast carrier relaxation dynamics$-$remain largely unexplored.

Excited-state processes are central to the performance of optoelectronic devices. Upon above-gap photoexcitation, hot electrons and holes are generated. These carriers rapidly dissipate excess energy via electron--phonon scattering as they relax toward the band edges, often preceding efficient charge extraction. Concurrently, nonradiative recombination provides an additional decay channel that depletes carrier populations. Understanding and controlling these competing processes is therefore essential for optimizing carrier lifetimes and charge transport.
At the microscopic level, these phenomena arise from the strong coupling between structural and electronic dynamics, driven by electron-phonon interactions and nonadiabatic (NA) transitions. Accurately describing the coupled electron--nuclear dynamics rqeuires methods that go beyond static electronic-structure calculations. \textit{Ab initio} nonadiabatic molecular dynamics (NAMD) combined with time-dependent density functional theory (TDDFT) \cite{runge1984density}, has emerged as a powerful framework for this purpose. These simulations have provided valuable insights into carrier relaxation in conventional perovskites \cite{zhang2026dimerization,ou2026cation}, demonstrating excellent agreement with ultrafast spectroscopic experiments \cite{wang2025self,panigrahi2022tailoring,wang2024detrimental,du2020crystal}.

In the present study, we address this gap through a comprehensive study of the electronic structure and carrier dynamics in the carbide antiperovskites Ca$_6$CSe$_4$ and Sr$_6$CSe$_4$. Using many-body perturbation theory (MBPT) \cite{reining2018gw,onida2002electronic}, we first provide an accurate quasiparticle description of the band structures and assess excitonic effects. We then employ TDDFT combined with NAMD to investigate HC cooling and nonradiative recombination. 
Our results show that A-site cation substitution strongly modulates lattice fluctuations, NA couplings, and electronic decoherence times, resulting in markedly different carrier lifetimes. In particular, enhanced decoherence and reduced NA couplings in Ca$_6$CSe$_4$ lead to significantly prolonged recombination times. In addition, both materials exhibit a pronounced HC cooling bottleneck near the band edges, with overall faster cooling in Ca$_6$CSe$_4$, arising from stronger intraband NA couplings. These findings offer microscopic insight into carrier relaxation mechanisms in antiperovskites and highlight the critical role of lattice-driven dynamics in governing their excited-state behavior.

\begin{figure*}[h]
	\centering
	\includegraphics[width=0.9\textwidth]{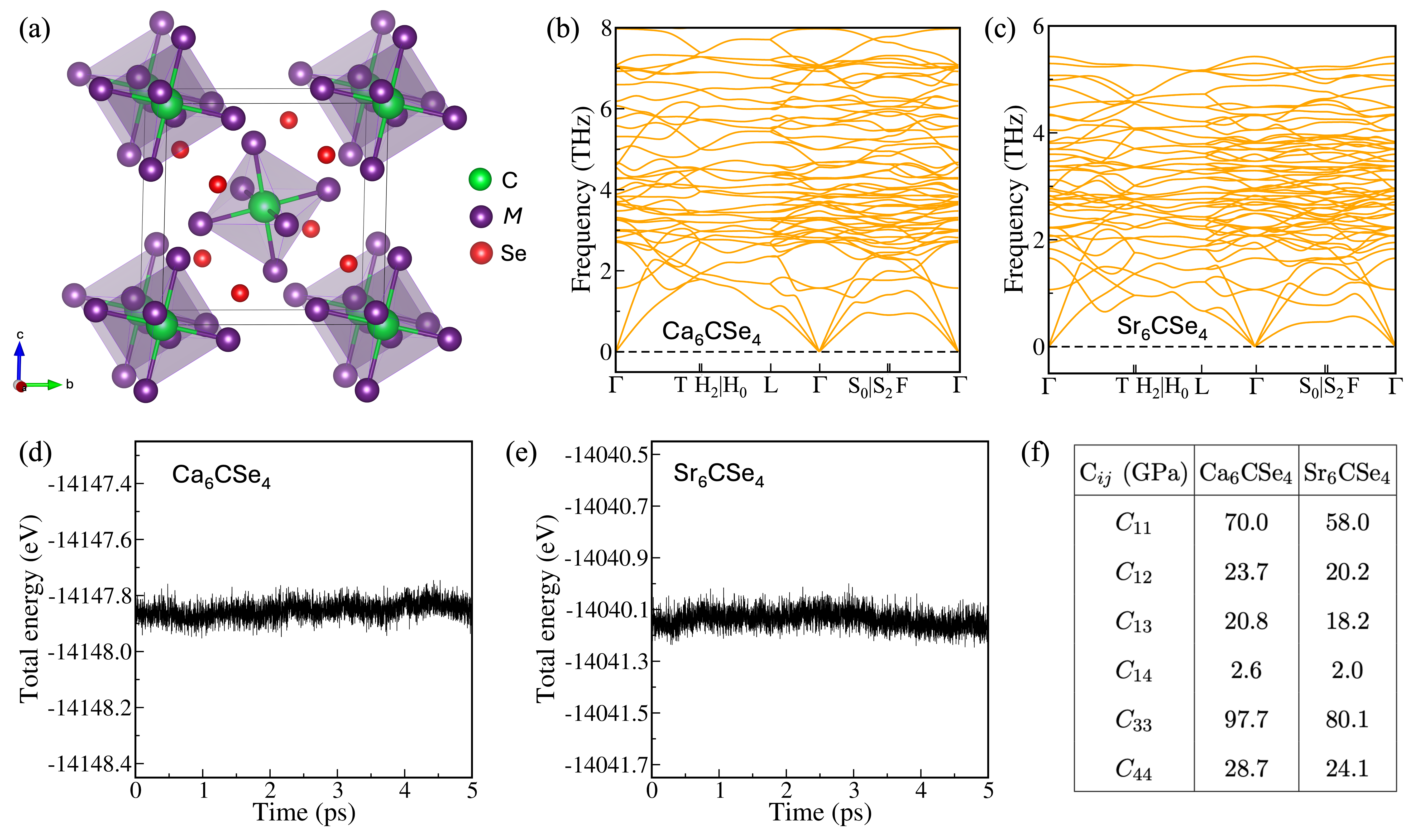}
	\caption{Structural and stability analysis of $M$$_6$CSe$_4$ ($M$ = Ca, Sr): (a) primitive unit cell; phonon dispersion for (b) Ca$_6$CSe$_4$ and (c) Sr$_6$CSe$_4$; total energy as a function of simulation time from $\textit{ab initio}$ molecular dynamics simulations in the canonical (NVT) ensemble at 300 K for (d) Ca$_6$CSe$_4$ and (e) Sr$_6$CSe$_4$; and (f) calculated elastic constants $C{ij}$ (in GPa), confirming mechanical stability.}
	\label{crys_stab}
\end{figure*}


The crystal structures of the carbide antiperovskites $M$$_6$CSe$_4$ ($M$ = Ca, Sr) are derived from zero-dimensional (0D) perovskite Cs$_4$Pb$X_6$, which adopts the $\mathit{R\overline{3}c}$ space group \cite{guo2022anti,tang2023promising}. In Cs$_4$Pb$X_6$, isolated Pb$X_6$ octahedra are separated by Cs cations. Through ion inversion, this structure is transformed into the antiperovskite $M$$_6$CSe$_4$, resulting in a significantly enhanced structural connectivity.
The atomic arrangement of $M$$_6$CSe$_4$ is illustrated in Fig.~\ref{crys_stab} (a), which shows the primitive unit cell, while the conventional cell is presented in Fig. S1 of the Supporting Information (SI). In contrast to the parent 0D structure, the $M$$_6$C octahedra in $M$$_6$CSe$_4$ are no longer isolated. The central C atoms form bonds with both the surrounding $M$-site cations (Ca or Sr) and interstitial Se atoms, as revealed by the electron localization function in Fig. S2 of SI. This enhanced connectivity increases the electronic dimensionality of the lattice, which is favorable for photoinduced charge transport. Substituting Ca with Sr at the $M$ site increases the intra-octahedral $M$–C bond length from 2.427 \text{\AA} to 2.603 \text{\AA}, and the inter-octahedral C–C distance from 7.151 \text{\AA} to 7.579 \text{\AA}. Consequently, the lattice constant of the primitive unit cell expands from 8.23 \text{\AA} in Ca$_6$CSe$_4$ to 8.71 \text{\AA} in Sr$_6$CSe$_4$, consistent with previous reports \cite{guo2022anti}.

To assess dynamical stability, we calculate the phonon dispersion plots, shown in Fig.~\ref{crys_stab} (b) and (c). The absence of imaginary frequencies throughout the Brillouin zone confirms that both Ca\(_6\)CSe\(_4\) and Sr\(_6\)CSe\(_4\) are dynamically stable. Thermal stability at room temperature is verified using AIMD simulations in the canonical (NVT) ensemble at 300~K (Fig.~\ref{crys_stab} (d) and (e)). The total energy remains nearly constant, and the temperature fluctuates narrowly around 300~K (Fig. S3 of SI), with no bond breaking observed, indicating robust thermal stability. 
Mechanical stability is evaluated by calculating the elastic constants. For Trigonal(I) systems, there are six independent elastic constants: \(C_{11}\), \(C_{12}\), \(C_{13}\), \(C_{14}\), \(C_{33}\), and \(C_{44}\). These satisfy the Born stability criteria \cite{mouhat2014necessary}:
\[C_{11} - |C_{12}| > 0\]
\[(C_{11} + C_{12}) C_{33} > 2 C_{13}^2\]
\[C_{44} > 0\]
\[(C_{11}-C_{12})C_{44}>2C_{14}^2\]
The calculated elastic constants (Fig.~\ref{crys_stab} (f)) satisfy all four conditions for both compounds, thereby confirming their mechanical stability.

\begin{figure}[h]
	\centering
	\includegraphics[width=0.7\textwidth]{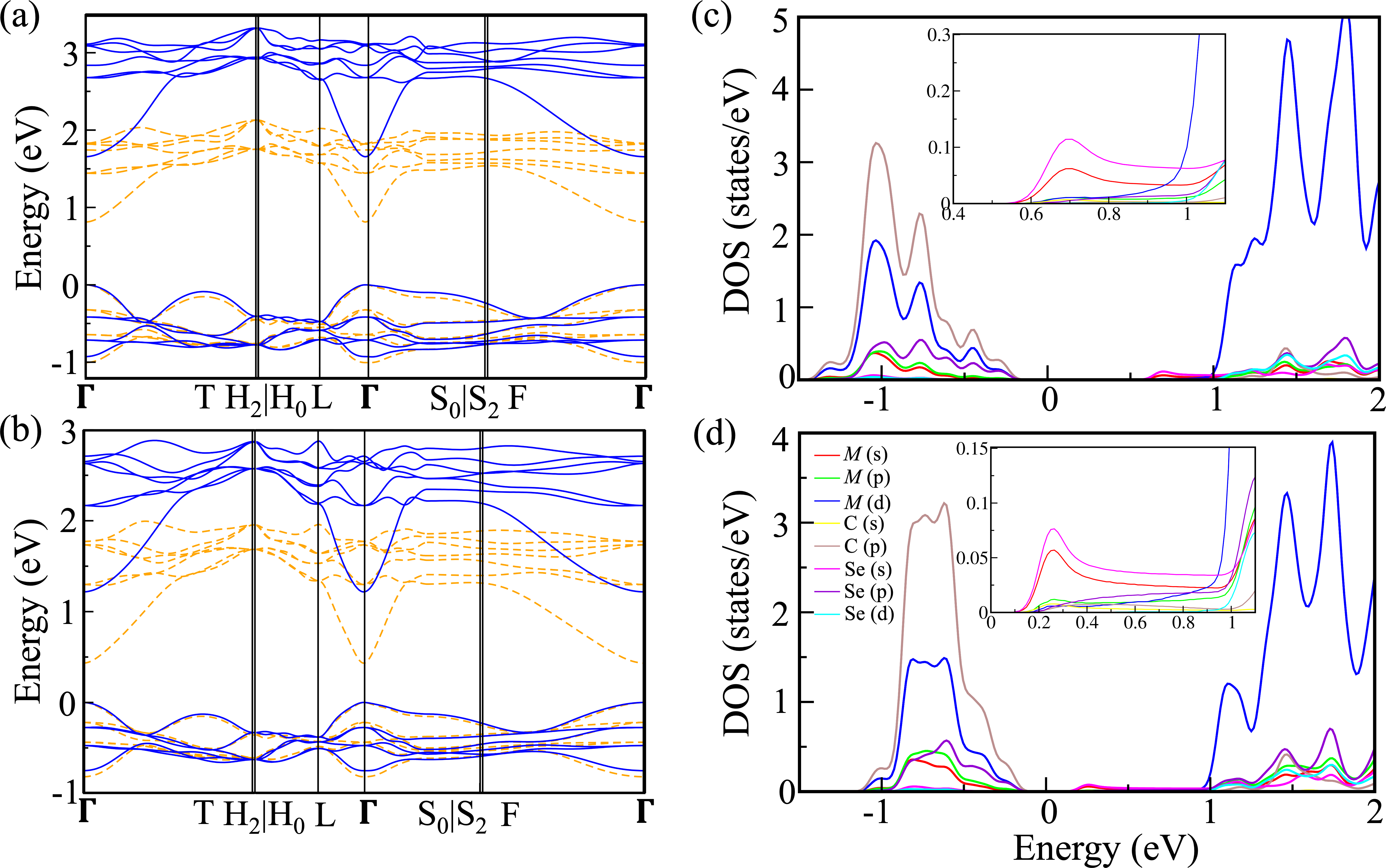}
	\caption{Electronic band structures computed using the PBE (orange dashed curves) and $\mathit{G_0W_0}$@PBE (blue solid curves) methods for (a) Ca$_6$CSe$_4$ and (b) Sr$_6$CSe$_4$.}
	\label{band_str_dos}
\end{figure}

We investigate the electronic structure of Ca$_6$CSe$_4$ and Sr$_6$CSe$_4$ using density functional theory (DFT) and many-body perturbation theory ($GW$). Figure~\ref{band_str_dos} shows the electronic band structures computed using the PBE (orange dashed lines) and single-shot $GW$ ($\mathit{G_0W_0}$@PBE) (blue solid lines) approaches for both compounds. At the PBE level, both materials exhibit direct band gaps at the $\Gamma$ point. Substituting Ca with Sr at the $M$ site substantially reduces the band gap, demonstrating that varying A-site cation provides an effective route for band-gap engineering in this family of antiperovskites. While the PBE functional captures the qualitative band dispersion, it is well known to underestimate band gaps. Calculations using the hybrid HSE06 functional give values of 1.39~eV for Ca$_6$CSe$_4$ and 1.03~eV for Sr$_6$CSe$_4$. A more accurate quasiparticle description is obtained within the $\mathit{G_0W_0}$ approximation, which opens the fundamental gaps to 1.66~eV and 1.22~eV, respectively (Table~\ref{bandgap}). The systematic increase in band gap from PBE to HSE06 to $\mathit{G_0W_0}$ highlights the importance of many-body corrections in these compounds. Notably, the $\mathit{G_0W_0}$ correction shifts the band edges while preserving their dispersion and nature of the band gap. Spin–orbit coupling effects are found to be negligible for these systems (see Table~S2 of SI) and are therefore not included in the present analysis.
\begin{table}[h]
	\centering
	\caption{Band gaps (in eV) of $M_6$CSe$_4$ ($M$ = Ca, Sr) calculated using different levels of theory. The exciton binding energy $E_b$ (eV) is estimated as the difference between the $\mathit{G_0W_0}$@PBE gap and the lowest bright excitation from the Bethe--Salpeter equation (BSE).}
	\label{bandgap}
	\begin{tabular}{|c|c|c|c|c|c|}
		\hline
		$M_6$CSe$_4$ & PBE & HSE06 & $\mathit{G_0W_0}$@PBE & BSE & $E_b$ \\
		\hline
		Ca$_6$CSe$_4$ & 0.79 & 1.39 & 1.66 & 1.54 & 0.12 \\
		Sr$_6$CSe$_4$  & 0.40 & 1.03 & 1.22 & 1.02 & 0.20 \\
		\hline
	\end{tabular}
\end{table}
Further insights into carrier transport is provided by the effective masses extracted from parabolic fitting near the band edges (Table~S3 of SI). The electron effective masses are smaller than the free-electron mass ($m_e^*/m_0 < 1$) and lower than the corresponding hole masses, suggesting high electron mobilities.

The orbital character of the band edges is revealed by the atom- and orbital-projected density of states (PDOS) shown in Fig.~\ref{band_str_dos} (c,d). The valence band maximum (VBM) is dominated by hybridized C-$p$ and Se-$p$ states with a noticeable contribution from $M$-$d$ orbitals. In contrast, the conduction band minimum (CBM) exhibits predominant $s$-character from the $M$ and Se atoms. The participation of formally empty $M$-$d$ states near the VBM arises from the small energy separation between the $ns$ and $(n-1)d$ levels in Ca and Sr, which enables partial $d$-occupation and facilitates $M$--C bonding \cite{guo2022anti}. This orbital mixing is consistent with observations in other antiperovskite systems \cite{mochizuki2020theoretical,sreedevi2022antiperovskite,han2021design} and plays an important role in shaping the band dispersion and optical transitions.

While quasiparticle calculations accurately determine the fundamental band gap, they do not account for electron–hole interactions that govern optical excitations. Solving the Bethe–Salpeter equation (BSE) on top of the $\mathit{G_0W_0}$ electronic structure yields exciton binding energies (E$_b$) of 120~meV for Ca$_6$CSe$_4$ and 200~meV for Sr$_6$CSe$_4$  (Table~\ref{bandgap}). The dielectric response and real-space exciton distributions is presented in Fig.~S5 of SI. The exciton wavefunctions extend over multiple unit cells, indicating a delocalized, Wannier–Mott-like character. Notably these values correspond to zero-temperature estimates. Under realistic conditions, lattice polarization, electron–phonon coupling, and dynamical screening are expected to further reduce the effective binding energies \cite{bokdam2016role,lizarraga2025determining}.

\begin{table}[h]
	\centering
	\caption{Root-mean-square fluctuation (RMSF) of atomic positions (in \AA) for the $M$-site, C, and Se atoms in Ca$_6$CSe$_4$ and Sr$_6$CSe$_4$, averaged over the last 5~ps of the AIMD trajectories.}
	\label{rmsf}
	\begin{tabular}{|c|c|c|c|}
		\hline
		$M_6$CSe$_4$ & $M$ & C & Se \\
		\hline
		Ca$_6$CSe$_4$ & 0.089 & 0.072 & 0.068 \\
		Sr$_6$CSe$_4$  & 0.069 & 0.072 & 0.068 \\
		\hline
	\end{tabular}
\end{table}
All preceding calculations are performed on optimized 0~K structures, providing a static description of the electronic and optical properties. At finite temperature, however, thermal fluctuations dynamically modulate the crystal lattice, which in turn perturbs the instantaneous electronic structure and influences carrier relaxation pathways. To capture these effects, we perform AIMD simulations at 300~K in the canonical (NVT) ensemble. Thermal excitation induces an expansion of both intra- and inter-octahedral bonds. The average $M$--C bond length increases to 2.429~\AA\ in Ca$_6$CSe$_4$ and 2.608~\AA\ in Sr$_6$CSe$_4$, while the inter-octahedral C--C distance expands to 7.188~\AA\ and 7.635~\AA, respectively. The larger expansion observed in Sr$_6$CSe$_4$ indicates its softer lattice. Thermal disorder is quantified by computing the root-mean-square fluctuation (RMSF) of atomic positions over the last 5~ps of the AIMD trajectories (Table~\ref{rmsf}). Substituting Ca with Sr significantly reduces the RMSF at the metal sites, reflecting the larger ionic radius and heavier mass of Sr, which suppress vibrational amplitudes. In contrast, the fluctuations of C and Se atoms remain nearly identical in both compounds. Overall, Sr$_6$CSe$_4$ exhibits smaller atomic displacements than Ca$_6$CSe$_4$, indicating greater structural rigidity at finite temperature.

\begin{figure}[h]
	\centering
	\includegraphics[width=0.43\textwidth]{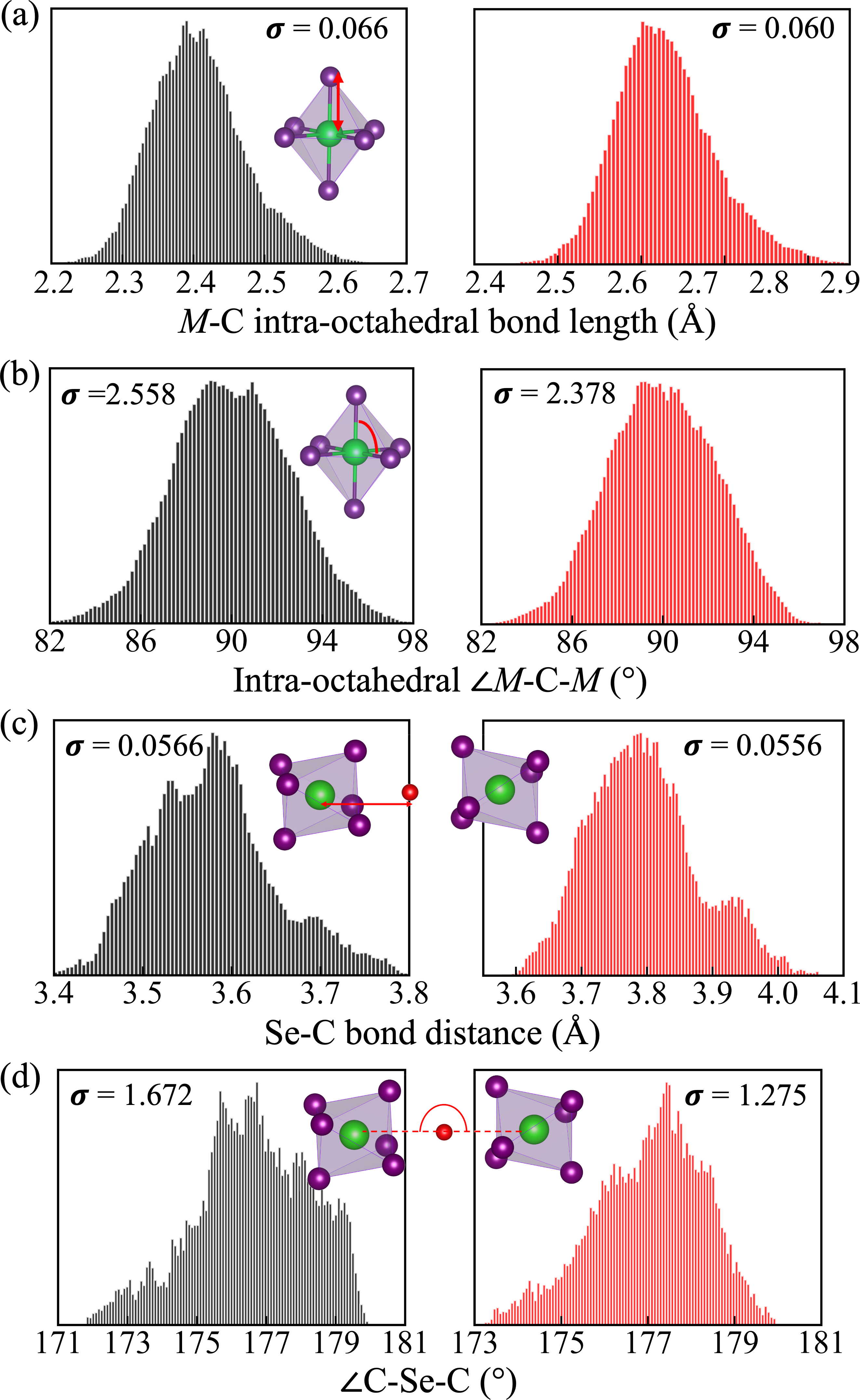}
	\caption{Statistical distributions of (a) intra-octahedral $M$--C bond lengths, (b) $\angle M$--C--$M$ bond angles, (c) Se--C bond distances, and (d) $\angle C$--Se--C bond angles obtained from the last 5~ps of the AIMD trajectories for Ca$_6$CSe$_4$ (black) and Sr$_6$CSe$_4$ (red). Standard deviations ($\sigma$) of the fluctuations are also reported.}
	\label{angle_bond_fluc}
\end{figure}
Deeper insight into the nature of these fluctuations is obtained from the statistical distributions of key structural parameters extracted from the last 5~ps of the trajectories (Fig.~\ref{angle_bond_fluc}). These include intra-octahedral $M$--C bond lengths, $\angle M$--C--$M$ bond angles, Se--C distances, and $\angle C$--Se--C bond angles. The bond angles $\angle$$M$–C–$M$ and $\angle$C–Se–C reflect distortions of the $M$$_6$C octahedra. The distributions are noticeably broader for Ca$_6$CSe$_4$ than for Sr$_6$CSe$_4$, with larger standard deviations, indicating stronger deviations from ideal octahedral geometry and a more dynamically disordered lattice.
\begin{figure}[h]
	\centering
	\includegraphics[width=0.52\textwidth]{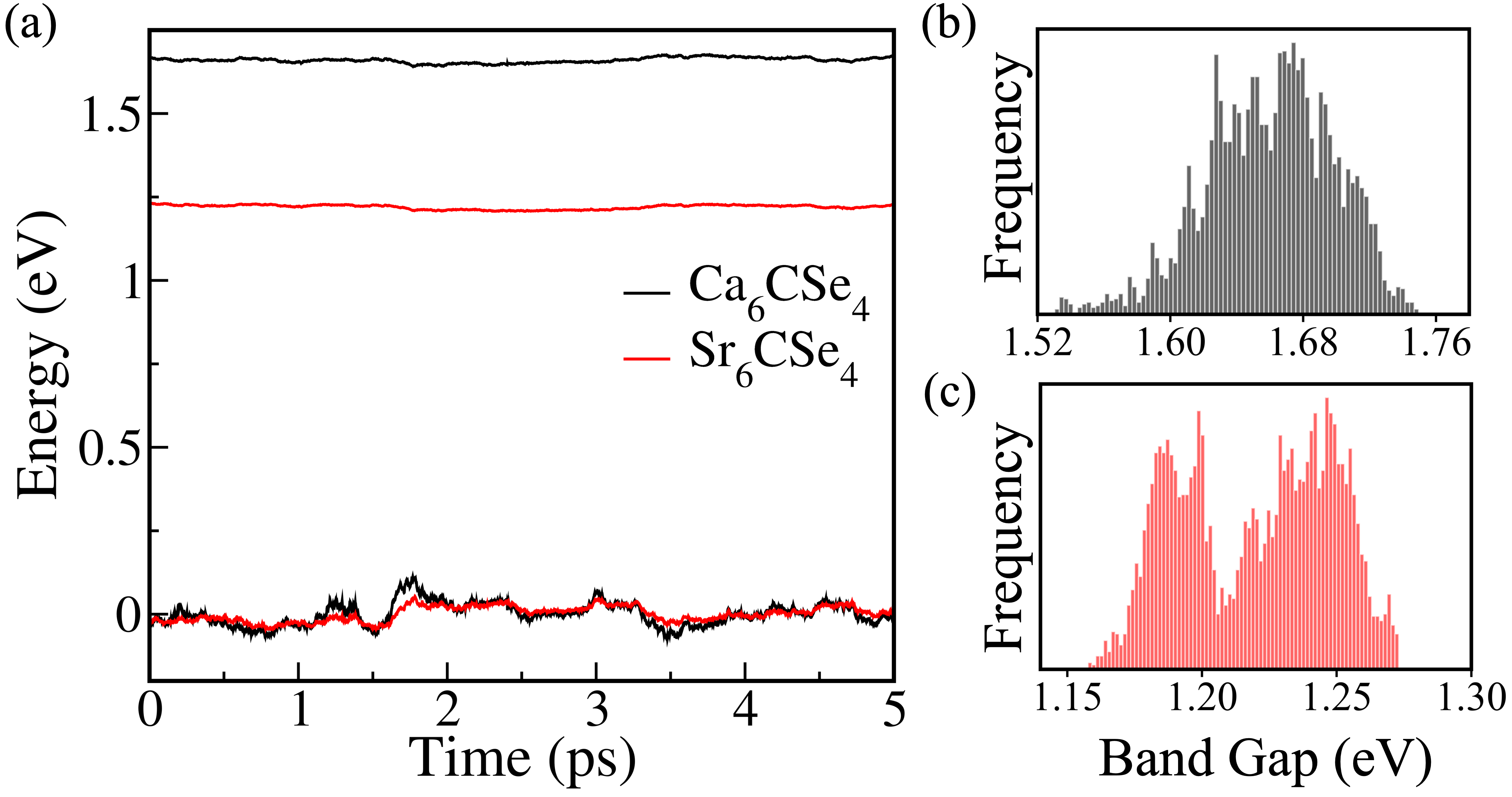}
	\caption{(a) Time evolution of the valence band maximum (VBM) and conduction band minimum (CBM) energies for Ca$_6$CSe$_4$ (black) and Sr$_6$CSe$_4$ (red). Panels (b) and (c) show the statistical distributions of the band gap for Ca$_6$CSe$_4$ and Sr$_6$CSe$_4$, respectively. The band gaps are scaled to the $\mathit{G_0W_0}$@PBE values. The standard deviation is larger for Ca$_6$CSe$_4$ ($\sigma = 0.0386$~eV) than for Sr$_6$CSe$_4$ ($\sigma = 0.0279$~eV).}
	\label{fluc}
\end{figure}
These structural fluctuations directly modulate orbital overlap and thereby perturb the instantaneous electronic structure. As shown in Fig.~\ref{fluc} (a), the VBM exhibits significantly larger temporal energy fluctuations than the CBM in both materials. This asymmetry originates from their distinct orbital characters: the VBM consists of directional C-$p$, Se-$p$, and $M$-$d$ states that are highly sensitive to bond-angle distortions, whereas the CBM is dominated by more isotropic $M$-$s$ and Se-$s$ states. Consequently, thermal lattice motion couples more strongly to the valence band edge. Consistent with its larger structural disorder, Ca$_6$CSe$_4$ displays stronger VBM fluctuations than Sr$_6$CSe$_4$. The enhanced fluctuations in Ca$_6$CSe$_4$ are also reflected in the band-gap distributions, depicted in Fig.~\ref{fluc} (b,c). The standard deviation of the band gap is larger for Ca$_6$CSe$_4$ ($\sigma = 0.0386$~eV) than for Sr$_6$CSe$_4$ ($\sigma = 0.0279$~eV). These dynamic band-edge fluctuations influence carrier localization and electronic coherence, thereby playing an important role in non-radiative recombination and HC relaxation dynamics.

\begin{figure*}[t]
	\centering
	\includegraphics[width=0.9\textwidth]{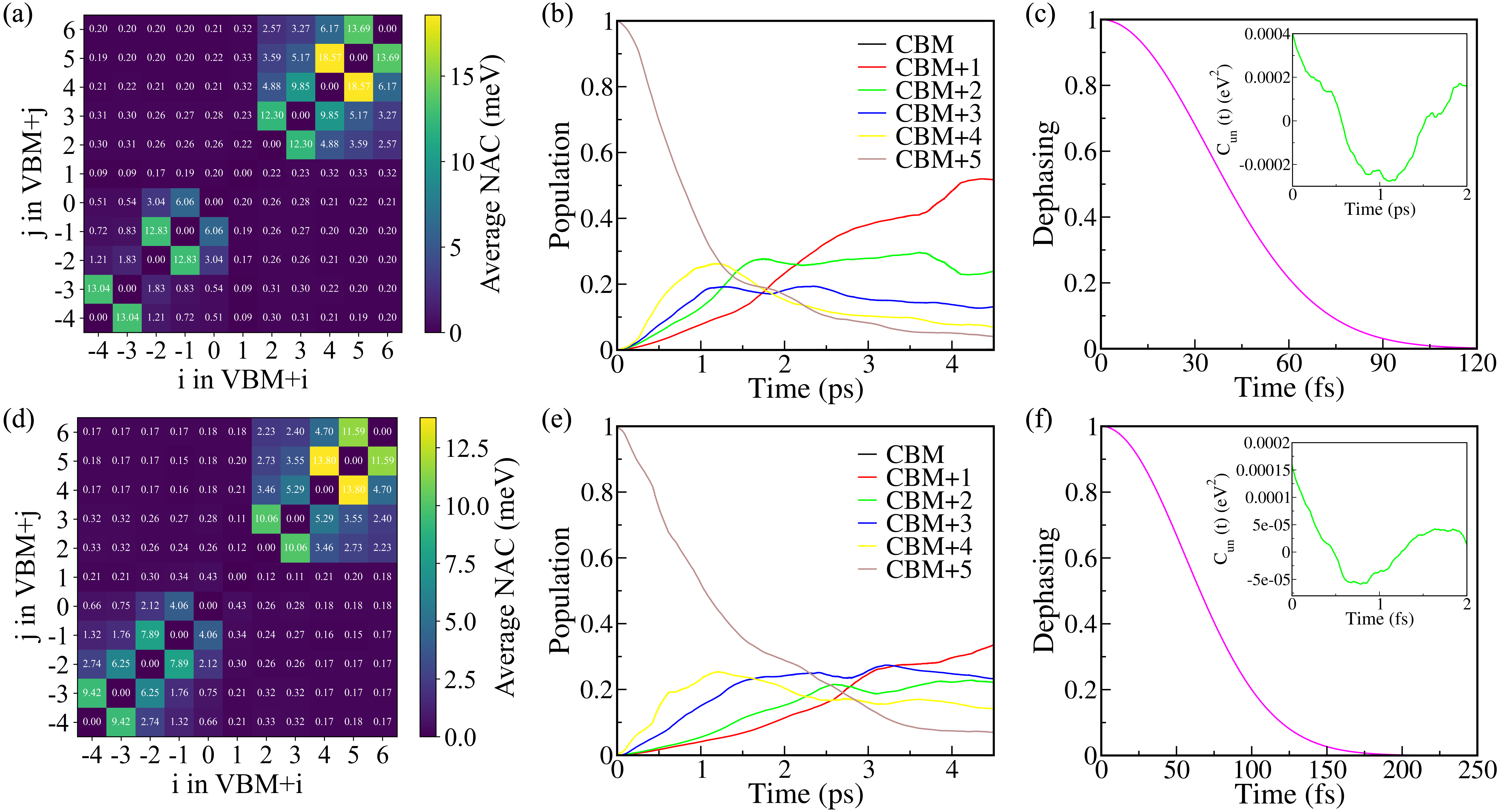}
	\caption{(a, d) Averaged absolute nonadiabatic (NA) couplings among states spanning VBM–4 to CBM+5 for Ca$_6$CSe$_4$ and Sr$_6$CSe$_4$, respectively. (b, e) Time evolution of hot-electron relaxation following excitation to CBM+5 for Ca$_6$CSe$_4$ and Sr$_6$CSe$_4$. (c, f) Pure-dephasing functions corresponding to the CBM+5 to CBM transition for Ca$_6$CSe$_4$ and Sr$_6$CSe$_4$. Insets display the unnormalized autocorrelation function, C$_{un}(t)$. The initial value, C$_{un}(0)$, represents the squared band-gap fluctuation amplitude; larger fluctuations lead to faster dephasing.}
	\label{hot_elec}
\end{figure*}

To probe hot-electron relaxation, we simulate photoexcitation from the VBM to the CBM+5 state, corresponding to excitation energies of 2.0–2.14~eV. The subsequent intraband relaxation of electrons within the conduction band is tracked from the CBM+5 state and shown in Fig.~\ref{hot_elec} (b,e). The decay of the excited-state population is fitted to an exponential function, $f(t)=\exp(-t/\tau)$, yielding cooling times of 1.1~ps for Ca$_6$CSe$_4$ and 1.6~ps for Sr$_6$CSe$_4$. 
As the electron approaches the band edge, the relaxation slows markedly (Table~S4 in SI), with particularly long lifetimes at the CBM+1 state (6.8~ps for Ca$_6$CSe$_4$ and 9.1~ps for Sr$_6$CSe$_4$). This bottleneck originates from the large energy separation between CBM+1 and CBM (Fig.~S6 of SI)$-$0.614~eV in Ca$_6$CSe$_4$ and 0.856~eV in Sr$_6$CSe$_4$$-$which suppresses phonon-assisted transitions.

To elucidate the microscopic origin of the cooling behavior, we analyze the role of electron–phonon interactions. Inelastic electron–phonon scattering governs energy dissipation by enabling transitions between electronic states via lattice vibrations. The strength of this interaction is quantified by the NA coupling:
\begin{equation}
d_{ij} = -i\hbar \langle \varphi_i | \frac{\partial}{\partial t} | \varphi_j \rangle 
= -i\hbar \frac{\langle \varphi_i | \nabla_R \hat{H} | \varphi_j \rangle}{\varepsilon_i - \varepsilon_j} \dot{R},
\label{nac}
\end{equation}
where $\varphi_{i/j}$ and $\varepsilon_{i/j}$ are the Kohn--Sham wavefunctions and eigenvalues, $\hat{H}$ is the Kohn--Sham Hamiltonian, and $\dot{R}$ is the nuclear velocity. In practice, $d_{ij}$ is evaluated from wavefunction overlaps along the AIMD trajectory, and its magnitude directly determines the probability of nonradiative transitions.

The averaged absolute NA coupling matrices, depicted as heat maps Fig.~\ref{hot_elec} (a,d), show that couplings between adjacent states are stronger than the nonadjacent states, confirming the dominant role of sequential intraband transitions in hot-electron relaxation. Nevertheless, finite off-diagonal elements indicate that non-adjacent jumps also contribute. Importantly, Ca$_6$CSe$_4$ exhibits systematically larger NA couplings across the conduction-band manifold compared to Sr$_6$CSe$_4$, which directly explains its faster cooling dynamics. This difference stems from the smaller energy separation between the conduction band states in Ca$_6$CSe$_4$  which enhance the denominator term in Eq.~\eqref{nac}. 
The slowing down of cooling rate at the CBM+1 state can also be attributed to the large energy separations (0.614 eV in Ca$_6$CSe$_4$ and 0.856 eV in Sr$_6$CSe$_4$) and correspondingly weak NA coupling (0.22 and 0.12 meV, respectively). This quantitative inverse scaling directly explains the observed HC cooling dynamics: rapid relaxation at high energies ($\sim$1 ps) and a pronounced slowdown near the band edge (up to $\sim$9 ps), where large energy gaps suppress phonon-assisted transitions.

In addition to inelastic scattering, elastic electron–phonon interactions also influence carrier dynamics by disrupting quantum coherence between electronic states. This decoherence is captured by the pure-dephasing function, obtained via the second-order cumulant expansion:
\begin{equation}
	D_{ij}(t) = \exp\left(-\frac{1}{\hbar^2} \int_0^t dt' \int_0^{t'} dt'' C_{ij}(t'') \right),
	\label{deph_eq}
\end{equation}
where $C_{ij}(t) = \langle \delta E_{ij}(t),\delta E_{ij}(0) \rangle$ is the unnormalized autocorrelation function of the phonon-induced energy-gap fluctuations between states $i$ and $j$, with C$_{un}(0)$ denoting its initial value. This quantity represents the squared amplitude of band-gap fluctuations, with larger values leading to faster dephasing.
 Here, $\delta E_{ij}(t) = E_{ij}(t) - \langle E_{ij} \rangle$ denotes the instantaneous deviation of the energy gap from its ensemble average. The pure-dephasing time is extracted by fitting $D_{ij}(t)$ to a Gaussian function of the form $\exp[-0.5(t/\tau)^2]$. As shown in Fig.~\ref{hot_elec} (c,f), Ca$_6$CSe$4$ exhibits a larger value of $C_{un}(0)$, indicating stronger energy-gap fluctuations and hence faster decoherence (34.2 fs for Ca$_6$CSe$_4$ and 55.8 fs for Sr$_6$CSe$_4$). Although faster dephasing generally reduces transition probabilities, the significantly stronger NA couplings in Ca$_6$CSe$_4$ dominate, resulting in overall faster hot-electron cooling.

\begin{figure*}[h]
	\centering
	\includegraphics[width=0.6\textwidth]{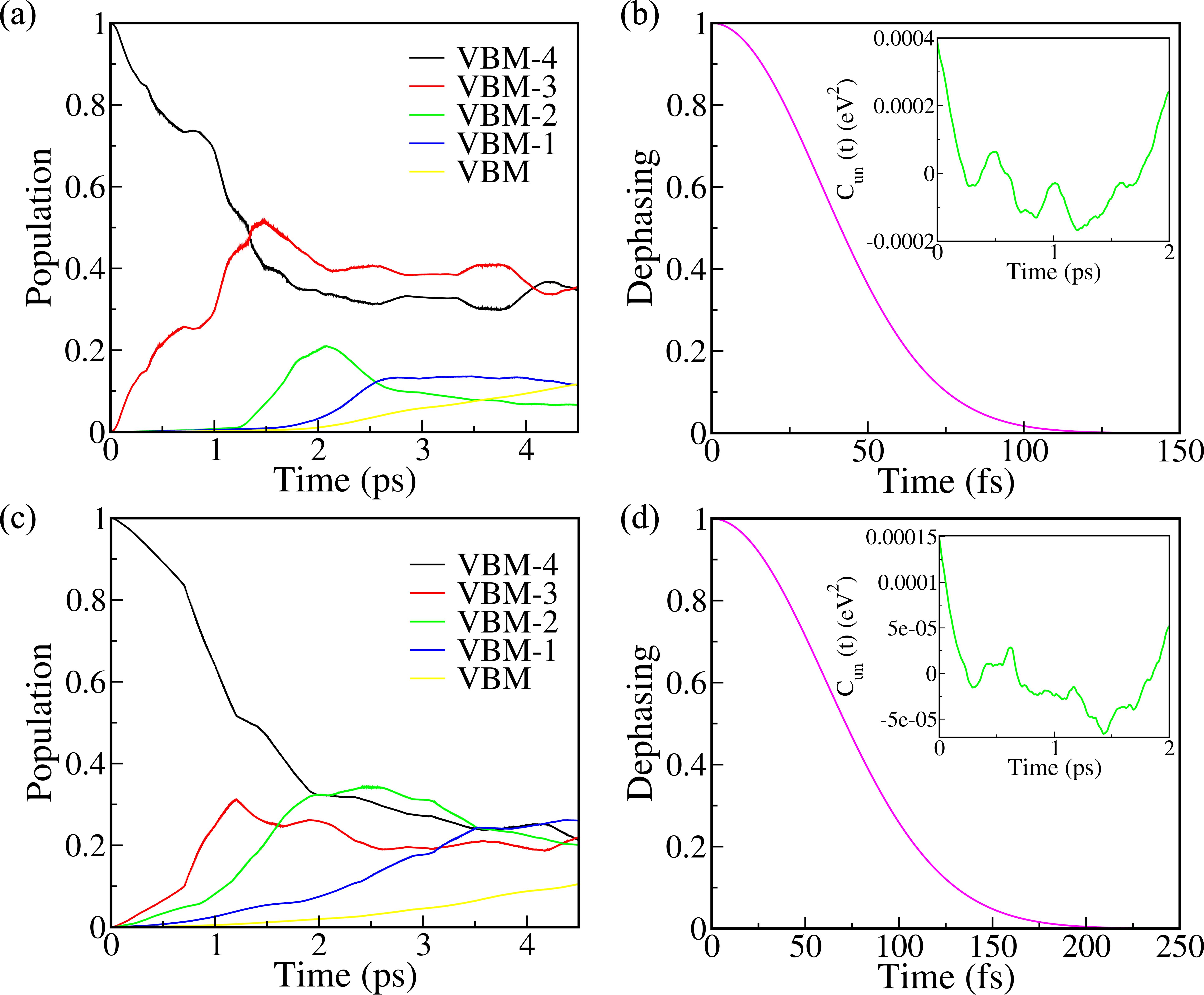}
	\caption{Hot-hole relaxation from VBM–4 to the VBM in (a) Ca$_6$CSe$_4$ and (c) Sr$_6$CSe$_4$. Panels (b) and (d) show the corresponding pure-dephasing functions for the VBM–4 to VBM transition in Ca$_6$CSe$_4$ and Sr$_6$CSe$_4$, respectively. The insets in (b) and (d) display the unnormalized autocorrelation functions of the VBM–4–VBM energy-gap fluctuations. A larger initial value, C$_{un}(0)$, indicates stronger gap fluctuations and generally corresponds to a shorter pure-dephasing time.}
	\label{hole_HC}
\end{figure*}

Hot-hole relaxation exhibits behavior qualitatively similar to hot-electron cooling. To investigate hot-hole dynamics, we promote an electron from VBM-4 to the CBM and monitor the evolution of hole populations within the valence-band manifold. The population decay and corresponding pure-dephasing functions are shown in Fig.~\ref{hole_HC}. Relaxation proceeds predominantly via sequential intraband transitions (VBM-4 $\rightarrow$ VBM-3 $\rightarrow$ VBM-2 $\rightarrow$ VBM-1 $\rightarrow$ VBM), consistent with the dominance of adjacent-state couplings observed in the NA matrices (Fig.~\ref{hot_elec}(a,d)).
The initial relaxation from VBM-4 exhibits comparable lifetimes in both compounds, 2.5 ps for Ca$_6$CSe$_4$ and 2.3 ps for Sr$_6$CSe$_4$. Thus occurs despite stronger NA coupling in Ca$_6$CSe$_4$, due to faster dephasing (35.1 fs in Ca vs 60.8 fs in Sr) which suppresses electronic transitions.
In Ca$_6$CSe$_4$, the larger energy separation between the VBM-2 and VBM-3 states (0.125 eV) reduces the NA coupling (1.83 meV), slowing relaxation at this step. In contrast, Sr$_6$CSe$_4$ has smaller energy spacings (0.038 eV) and stronger NA coupling (6.25 meV), leading to faster relaxation through these states.
Closer to the band edge, a pronounced slowdown occurs at VBM-1, with lifetimes of 4.2 ps (Ca$_6$CSe$_4$) and 6.8 ps (Sr$_6$CSe$_4$). This slowdown arises from large energy separation and reduced NA couplings, which are weaker in Sr$_6$CSe$_4$ (4.06 meV vs 6.06 meV in Ca), leading to a longer lifetime at VBM-1 and transient hole accumulation.

Overall, both hot-electron and hot-hole cooling in $M_6$CSe$_4$ occur on the picosecond timescale (1--9~ps), with a clear slowdown near the band edges. This behavior contrasts with much faster HC cooling in conventional semiconductors, such as $\sim$50~fs in GaAs, $\sim$0.4~ps in non-nanostructured Si, and $\sim$2--3~ps in CIGS \cite{bernardi2015ab,laurell2025coherent,bernardi2014ab,doany1988measurement}. In lead halide perovskites such as MAPbI$_3$, cooling times vary widely depending on excitation conditions and sample morphology, ranging from sub-picosecond relaxation in bulk films to tens or hundreds of picoseconds (and even nanoseconds) under specific conditions \cite{man2026towards}. These results demonstrate that HC relaxation in $M_6$CSe$_4$ can be systematically tuned through composition-driven modifications of the electronic structure and lattice dynamics.

\begin{figure}[h]
	\centering
	\includegraphics[width=0.7\textwidth]{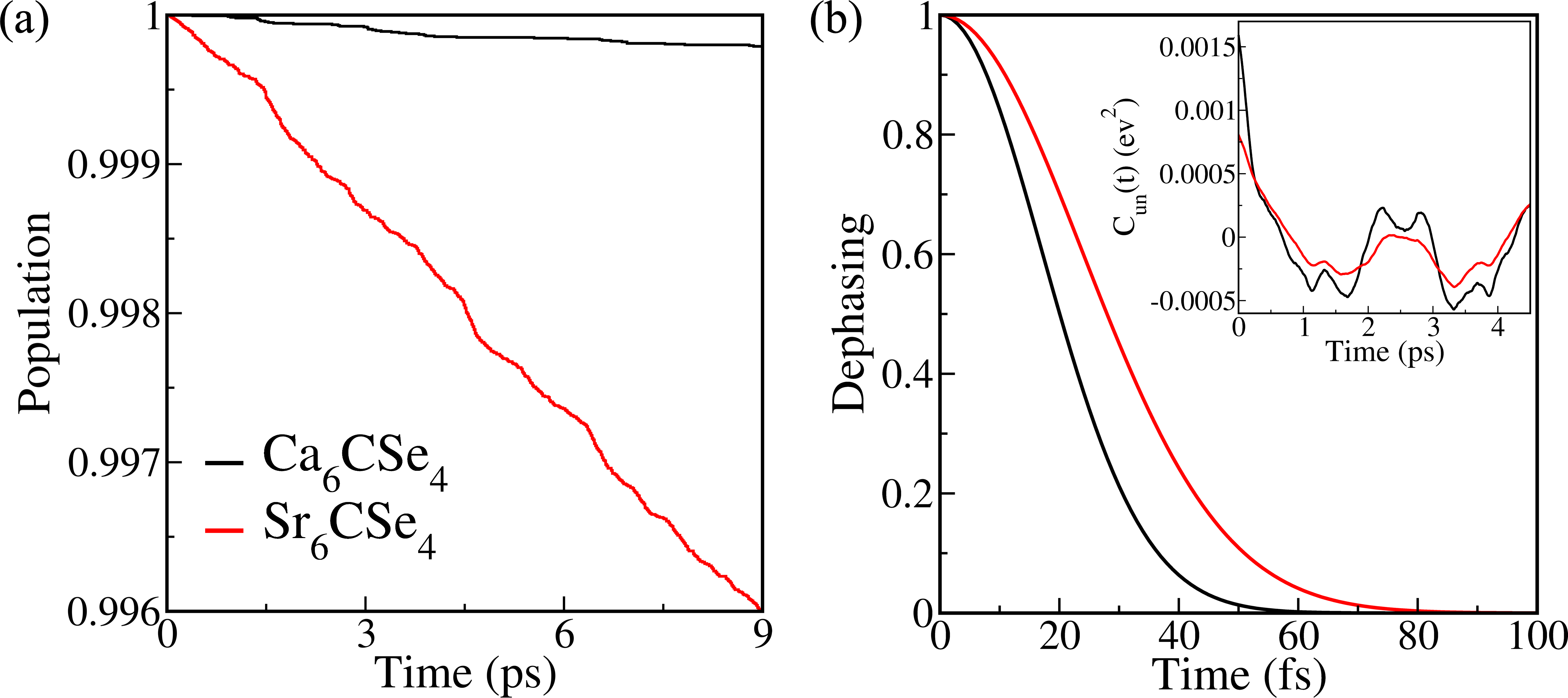}
	\caption{Time evolution of the population of the first excited state corresponding to the VBM--CBM transition.}
	\label{recomb_deph}
\end{figure}
Figure~\ref{recomb_deph} (a) shows the time evolution of the excited-state population associated with the VBM--CBM transition in $M_6$CSe$_4$. Because intraband HC cooling occurs on picosecond timescales---much faster than recombination---we assume that electrons and holes have relaxed to the CBM and VBM, respectively, before recombination takes place. The decay of the excited-state population is fitted with a single exponential function $\exp(-t/\tau)$, yielding non-radiative recombination lifetimes of 40.3~ns for Ca$_6$CSe$_4$ and 2.2~ns for Sr$_6$CSe$_4$ (Table~\ref{lifetimes}). The nearly order-of-magnitude difference highlights a strong suppression of non-radiative recombination in the Ca-based compound. The recombination lifetimes are governed by the interplay of three key factors: the band-gap magnitude, the strength of NA couplings, and the electronic decoherence time \cite{ma2024compression,dai2022improving}. To obtain quantitatively reliable estimates, the time-averaged PBE band gaps are rescaled to the $\mathit{G_0W_0}$@PBE values.
\begin{table}[h]
	\centering
	    \caption{Canonically averaged scaled band gaps (in eV), with original band gaps included in parenthesis, averaged absolute NA coupling ($\langle |d_{ij}| \rangle$), root mean square (RMS) NA coupling ($d_{ij}^{\mathrm{RMS}}$), pure-dephasing time ($\tau_{\phi}$), and nonradiative recombination lifetime ($\tau_{\mathrm{nr}}$) for the VBM--CBM transition.}
    \label{lifetimes}
\begin{tabular}{|l c c c c c c c c c c|}
	\hline
	$M$$_6$CSe$_4$ && Gap && $\langle |d_{ij}| \rangle$ && $d_{ij}^{\mathrm{RMS}}$ && $\tau_{\phi}$ && $\tau_{\mathrm{nr}}$ \\
	&& (eV) && (meV) && (meV) && (fs) && (ns) \\
	\hline
	Ca$_6$CSe$_4$ && 1.66 (0.77) && 0.20 && 0.34 && 17.0 && 40.3 \\
	\hline
	Sr$_6$CSe$_4$ && 1.22 (0.37) && 0.43 && 0.76 && 23.7 && 2.2 \\
	\hline
\end{tabular}
\end{table}

First, Sr$_6$CSe$_4$ exhibits significantly stronger NA couplings---more than twice those of Ca$_6$CSe$_4$, thereby accelerating nonradiative recombination. Notably, the stronger coupling in Sr$_6$CSe$_4$ occurs despite its reduced structural fluctuations. This counterintuitive result is explained by its smaller band gap, which reduces the energy denominator in the NA coupling expression (Eq.~\eqref{nac}) and thereby amplifies $d_{ij}$.
Second, electronic decoherence modulates the recombination rate. The pure-dephasing times extracted from Fig.~\ref{recomb_deph} (b) are 17.0~fs for Ca$_6$CSe$_4$ and 23.7~fs for Sr$_6$CSe$_4$. Faster decoherence in Ca$_6$CSe$_4$, driven by its larger band-gap fluctuations (Fig.~\ref{fluc}), suppresses electronic transitions between the VBM and CBM, thereby reducing the recombination rate. This behavior is consistent with the higher initial value of the unnormalized autocorrelation function, $C_{\mathrm{un}}(0)$ in Ca$_6$CSe$_4$.
Collectively, the larger band gap, weaker NA couplings, and faster decoherence in Ca$_6$CSe$_4$ act cooperatively to inhibit non-radiative recombination, resulting in its substantially longer carrier lifetime. In contrast, Sr$_6$CSe$_4$ combines a smaller band gap, stronger NA couplings, and slower decoherence, all of which favor faster recombination.

The predicted lifetimes can be placed in the broader context of hybrid halide perovskites such as MAPbI$_3$, which typically exhibit intrinsic theoretical lifetimes in the range of $\sim$0.8–3~ns, extending beyond 10~ns when structural effects such as twin boundaries or nuclear quantum effects are considered \cite{chen2018communicating,tian2025extending,liu2025unraveling,liu2023nuclear}. Experimentally, carrier lifetimes in MAPbI$_3$ have been further enhanced from the nanosecond to the microsecond regime through strategies such as defect passivation and interface engineering \cite{gong2016electron}.
In this context, the intrinsic lifetimes of the present compounds$-$particularly Ca$_6$CSe$_4$$-$ indicates intrinsically suppressed nonradiative recombination, highlighting their potential as promising candidates for high-efficiency photovoltaic applications.

In summary, we present a comprehensive first-principles investigation of the electronic structure and excited-state carrier dynamics in the carbide antiperovskites Ca$_6$CSe$_4$ and Sr$_6$CSe$_4$. Our $\mathit{G_0W_0}$–BSE calculations reveal direct band gaps in the visible-to-infrared range for both compounds, along with moderate exciton binding energies, providing a reliable description of their optical response.
NAMD simulations further elucidate the key factors governing carrier dynamics. HC cooling occurs on the picosecond timescale and exhibits pronounced band-edge bottlenecks arising from large energy level separations and weak NA couplings, with faster cooling in Ca$_6$CSe$_4$. This intrinsic slowing of carrier relaxation, compared to conventional semiconductors, suggests potential for HC retention. In parallel, nonradiative recombination lifetimes are dictated by a delicate interplay between band gap, NA coupling strength, and electronic decoherence. Notably, Ca$_6$CSe$_4$ exhibits a significantly longer lifetime of 40.3~ns—nearly an order of magnitude higher than the 2.2~ns obtained for Sr$_6$CSe$_4$. This contrasting behavior between cooling and recombination reflects the combined influence of band-gap magnitude, NA coupling, and decoherence dynamics.

These findings identify A-site cation substitution as an effective strategy to tune the electronic structure, lattice fluctuations, and electron–phonon interactions, thereby enabling control over excited-state dynamics. The coexistence of picosecond-scale cooling, pronounced band-edge bottlenecks, and nanosecond carrier lifetimes highlights carbide antiperovskites as promising lead-free materials for optoelectronic and photovoltaic applications. These results motivate future experimental efforts aimed at fully realizing the potential of these materials. More broadly, this work establishes composition-driven lattice engineering as a powerful framework for tailoring excited-state dynamics in antiperovskites and related material systems.
\section{Associated Content}
\textbf{Supporting Information}

Computational Methods; Conventional cell of $M_6$CSe$_4$ ($M$ = Ca, Sr); Electron localization function; Temperature evolution along the AIMD trajectory; Charge density distributions corresponding to the VBM and CBM; Imaginary part of the dielectric function obtained from $\mathit{G_0W_0}$@PBE and BSE@$\mathit{G_0W_0}$@PBE approaches, along with the spatial distribution of the first bright exciton.; Time evolution of the electronic energy levels (from VBM–4 to CBM+5) along the NAMD trajectory; Summary of computational parameters employed in the YAMBO package; Lowest band gaps of antiperovskites computed using the PBE functional, without and with SOC; Effective masses of electrons and holes at the band edges; Hot-carrier relaxation lifetimes corresponding to the VBM+$i$ (hot-hole relaxation) and CBM+$i$ states (hot-electron relaxation).

\section{Author Information}

\textbf{Corresponding Authors}

\textbf{Sanchi Monga}$-$Department of Physics, Indian Institute of Technology Delhi, New Delhi 110016, India, Email: sanchi@physics.iitd.ac.in

\textbf{Saswata Bhattacharya}$-$Department of Physics, Indian Institute of Technology Delhi, New Delhi 110016, India, Email: saswata@physics.iitd.ac.in\\

\noindent \textbf{Author Contributions}

All authors contributed equally to this work.\\

\noindent \textbf{Notes}

The authors declare no competing financial interest.

\section{Acknowledgment}
S.M. acknowledge IIT Delhi for the senior research fellowship. S.B. acknowledge financial support from SERB under a core research grant [Grant no. CRG/2019/000647] to set up his High Performance Computing (HPC) facility ``Veena'' at IIT Delhi for computational resources.

\bibliography{references}

\end{document}